\documentclass[12pt,aps,prd,reprint,sort&compress, superscriptaddress, showpacs, nofootinbib, preprintnumbers]{revtex4-1}

\usepackage{graphicx}
\usepackage{amsmath}
\usepackage{amssymb}
\usepackage{dsfont}
\usepackage{amsfonts}
\usepackage[UKenglish]{babel}
\usepackage{hyperref}
\usepackage{nicefrac}
\usepackage{verbatim}
\usepackage{bbm}
\usepackage{color}
\usepackage{multirow}
\usepackage{subcaption}
\usepackage{cleveref}

\newcommand{\E}{\mathrm{e}}

\begin{document}

\title{Evidence for a new $SU(4)$ symmetry with $J=2$ mesons}
 
\author{M.~Denissenya}
\email{mikhail.denissenya@uni-graz.at}
\author{L.~Ya.~Glozman}
\email{leonid.glozman@uni-graz.at}
\author{M.~Pak}
\email{markus.pak@uni-graz.at}
\affiliation{Institut f\"ur Physik, FB Theoretische Physik, Universit\"at Graz, Universit\"atsplatz 5,
8010 Graz, Austria}

\begin{abstract}
Recently, a new symmetry of mesons has been found upon truncation
of the quasi-zero modes of the Overlap Dirac operator in lattice simulations.
Namely, the $\rho,\rho',\omega,\omega',a_1, b_1,h_1$ and possibly $f_1$
$J=1$ mesons get degenerate after removal of the quasi-zero modes. This
emergent symmetry has been established to be 
$SU(4)\supset SU(2)_{\textsc{L}} \times SU(2)_{\textsc{R}} \times U(1)_{\textsc{A}}$. It is higher
than the symmetry of the QCD Lagrangian and provides not only a
mixing of quarks of  given chirality in the isospin space, but
also the mixing of left-handed and right-handed components. Here we study, with the
Overlap Dirac operator, the isovector $J=2$ mesons upon the quasi-zero mode reduction
and observe a similar degeneracy. This result
further supports the $SU(4)$ symmetry in mesons of given spin
$J \geq 1$.  
\end{abstract}

\maketitle

\section{Introduction}
In recent $N_{\textsc{f}}=2$ dynamical lattice simulations with the Overlap Dirac operator
a large degeneracy of the spin $J=1$ $\rho,\rho',\omega,\omega',a_1, b_1,h_1$ and possibly $f_1$ mesons has been discovered upon removal of the lowest-lying
eigenmodes of the Dirac operator from the valence quark propagators \cite{Denissenya:2014poa, Denissenya:2014ywa}\footnote{A hint for this symmetry has been seen earlier
with the Chirally Improved Dirac operator \cite{Glozman:2012fj}; for a previous
lattice study on the low-mode truncation see Ref.~\cite{Lang:2011vw}.}. The correlators have demonstrated a clean exponential decay suggesting that
the mesons survive this truncation. 

One expects a priori, that upon elimination
of the quasi-zero modes of the Dirac operator, the chiral symmetry should be
restored, since the quark condensate of the vacuum is connected with
the density of the quasi-zero modes via the Banks-Casher relation \cite{Banks:1979yr}.
However, it has turned out that not only  degeneracy patterns from
the $SU(2)_{\textsc{L}} \times SU(2)_{\textsc{R}}$ and $U(1)_{\textsc{A}}$ symmetries are observed, but
 a larger degeneracy  that includes all possible
chiral multiplets for $J=1$ mesons. This symmetry  has been established
to be  $SU(4)\supset SU(2)_L \times SU(2)_R \times U(1)_{\textsc{A}}$ that includes both the isospin rotations of 
quarks of  given chirality as well as the rotations of chirality itself (chiralspin rotations) \cite{Glozman:2014mka, Glozman:2015qva}. 

The $SU(4)$ symmetry is higher than the $SU(2)_{\textsc{L}} \times SU(2)_{\textsc{R}} \times U(1)_{\textsc{A}}$
symmetry of the QCD Lagrangian and should be consequently considered as an emergent symmetry
that reflects the QCD dynamics in $J=1$ mesons without the quasi-zero modes
of the Dirac operator. This symmetry implies the absence of the color-magnetic
field in the system and might be interpreted as a manifestation of the dynamical QCD string.

In the present paper we study the isovector $J=2$ mesons upon the reduction
of the lowest-lying eigenmodes from the Overlap valence quark propagators. We obtain the
same symmetry patterns as for $J=1$ mesons, which supports consequently
the existence of the $SU(4)$ symmetry in $J \geq 1$ mesons after the quasi-zero
mode reduction.

The structure of the article is as follows: In Chapter \ref{Chapter-Chiral-Parity} we discuss
the parity-chiral, chiralspin and $SU(4)$ multiplets for the tensor
mesons and respective interpolators and our expectations for the truncation of
the lowest-lying Dirac eigenmodes. In Chapter \ref{Chapter-Lattice-Setup} we describe
the lattice technicalities. 
The results are presented in Chapter \ref{Chapter-Results}. Conclusions are given
in Chapter \ref{Conclusions}. Additional tables are provided in Appendix \ref{Chapter-Appendix}.  
 
\section{Theoretical predictions for the low-mode removal}
\label{Chapter-Chiral-Parity}
We discuss the theoretical predictions for the tensor meson spectrum after low-mode removal, which come either from the 
$SU(2)_{\textsc{L}} \times SU(2)_{\textsc{R}}$ and $U(1)_{\textsc{A}}$ restorations Refs.~\cite{Glozman:2003bt,Glozman:2007ek},
or from the higher  $SU(4)$ symmetry, Refs.~\cite{Glozman:2014mka, Glozman:2015qva}. We also present
interpolators with the respective symmetry transformation properties.

\subsection{Chiral symmetry predictions}
\begin{figure}
\centering
\includegraphics[angle=0,width=.85\linewidth]{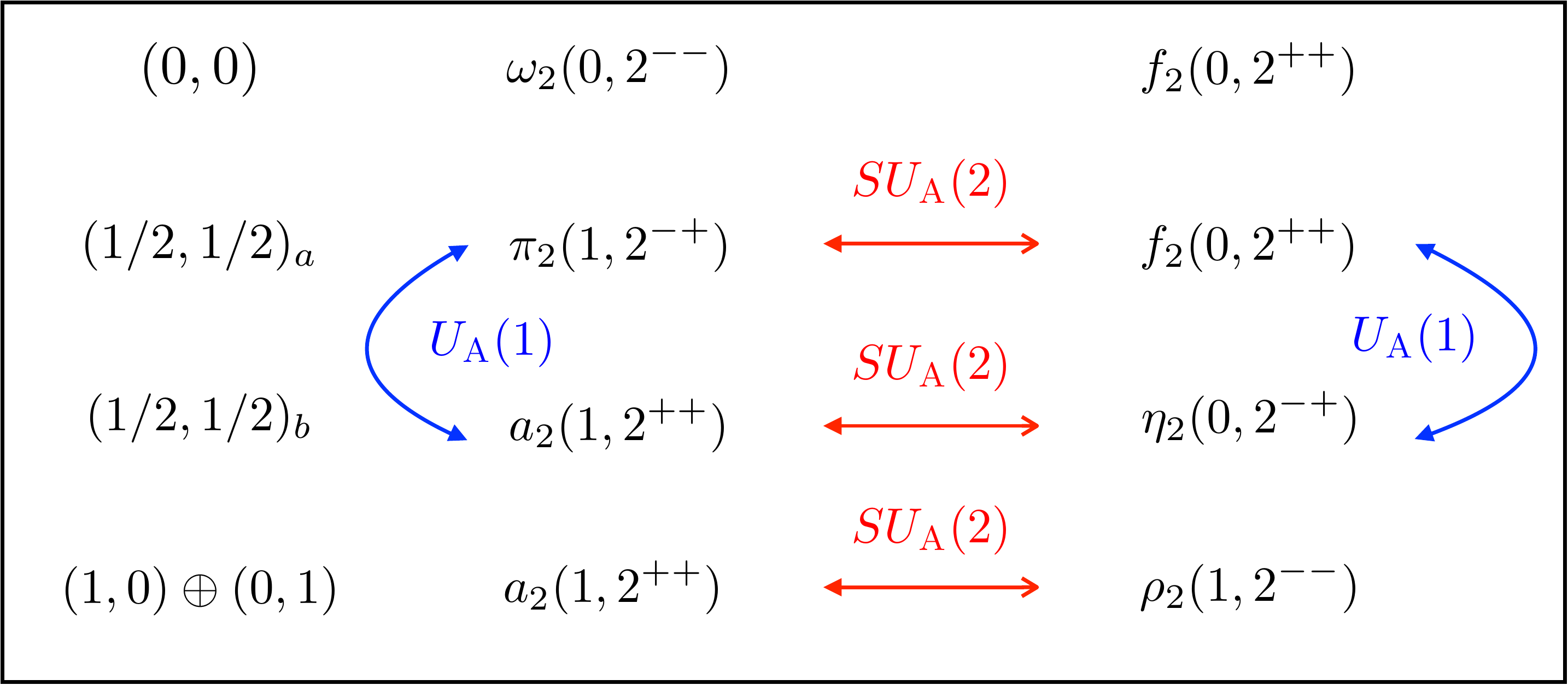}
\caption[Chiral-Parity group 1]{\sl On the left row the irreducible representations $r$ of $SU(2)_{\textsc{L}} \times SU(2)_{\textsc{R}}
\times C_i$ are given. 
Each meson is denoted as $(I, J^{PC})$, with $I$ isospin, $J$ total angular momentum, $P$ parity and $C$ charge conjugation.
The $SU(2)_{\textsc{A}}$ and $U(1)_{\textsc{A}}$ connections are denoted by red and blue lines, respectively.}
 \label{Table1}
\end{figure}

In Fig.~\ref{Table1} we show the classification of tensor mesons according to 
the parity-chiral group $SU(2)_{\textsc{L}} \times SU(2)_{\textsc{R}} \times \mathcal{C}_i$,
with $\mathcal{C}_i$ denoting the parity group.  

In a situation, where the $SU(2)_{\textsc{L}} \times SU(2)_{\textsc{R}}$
chiral symmetry is restored but no higher symmetry is present,
only  mesons within each parity-chiral multiplet $r$ must be degenerate.
For instance, in $(1,0) \oplus (0,1)$ the $a_2$ 
and $\rho_2$ states, which are related via $SU(2)_{\textsc{A}}$ should have the same mass\footnote{$SU(2)_{\textsc{A}}$ is  the shorthand notation for
the axial part of the $SU(2)_{\textsc{L}} \times SU(2)_{\textsc{R}}$ transformations.}.
If $U(1)_{\textsc{A}}$ is restored as well, all four mesons in the $(1/2,1/2)_a$ and
$(1/2,1/2)_b$ representations should be mass degenerate, {\it i.e.} the 
$SU(2)_{\textsc{L}} \times SU(2)_{\textsc{R}} \times U(1)_{\textsc{A}}$ restoration requires the following degeneracies
\footnote{The $a_2$ meson in $(1/2,1/2)_b$ is here denoted as $a'_2$, to distinguish it from the $a_2$ meson in $(1,0)\oplus (0,1)$. The same is 
true for the $f_2$ meson, which in $(1/2,1/2)_a$ is denoted as $f'_2$ to distinguish it from the $f_2$ in $(0,0)$.}
\begin{align}
\label{states-1}
 \pi_2 \longleftrightarrow f_2 \longleftrightarrow a'_2 \longleftrightarrow \eta_2 \; ,
\end{align}
and 
\begin{align}
\label{states-2}
a_2 \longleftrightarrow \rho_2 \; . 
\end{align}

\noindent The states from the singlet $(0,0)$ representations are invariant
with respect to both $SU(2)_{\textsc{L}} \times SU(2)_{\textsc{R}}$ and $U(1)_{\textsc{A}}$ transformations and consequently these symmetries do not constrain
their masses. 

One of the most interesting features of the parity-chiral group, Fig.~1, is, that two independent $2^{++}$
isovector, $a_2, a'_2$, and two independent $2^{++}$ isoscalar, $f_2, f'_2$, mesons must exist.
They differ from each other by the content of left- and right-handed quarks and therefore 
by the chiral representation $r$. They are not connected via a $SU(2)_{\textsc{L}} \times SU(2)_{\textsc{R}}$ or $U(1)_{\textsc{A}}$ transformation, 
so their masses should be different without additional symmetry constraints.
If their masses are degenerate, then there is a symmetry connecting the states of Eq.~(\ref{states-1}) with the states of Eq.~(\ref{states-2}). 
Hence, in such a situation a larger symmetry than $SU(2)_{\textsc{L}} \times SU(2)_{\textsc{R}} \times U(1)_{\textsc{A}}$
has to be present. This issue will be discussed in the next subsection.

In Table \ref{Table2} we classify the interpolators $\mathcal{O}_i$ used in our lattice study into the irreducible representations $r$
of the parity-chiral group.

\begin{table*}[thb]
\begin{center}
\begin{tabular}{|c|c|c|c|}
 \hline
 \hline
 \multicolumn{1}{|c|}{   $I,J^{PC}$ } &\multicolumn{1}{c|}{   $\mathcal{O}$  }  & $r$  
 &$O_h$  
 \\\hline
 
\multicolumn{1}{|c|}{\multirow{4}{*}{$\rho_2(1,2^{--}$)}} & $Q_{ijk} \bar{a}_{\partial_k}
\gamma_j\gamma_5 b_n-Q_{ijk} \bar{a}_n \gamma_j\gamma_5 b_{\partial_k}$ 
 & \multicolumn{1}{c|}{\multirow{2}{*}{ $(1,0)\oplus(0,1)$}} & \multicolumn{1}{c|}{\multirow{2}{*}{$E$}}
\\
 & $Q_{ijk} \bar{a}_{\partial_k} \gamma_j\gamma_5 b_w-Q_{ijk} \bar{a}_w \gamma_j\gamma_5
b_{\partial_k}$ 
 &  & 
 \\\cline{2-4}
  & $|\epsilon_{ijk}| \bar{a}_{\partial_k} \gamma_j\gamma_5 b_n-|\epsilon_{ijk}| \bar{a}_n
\gamma_j\gamma_5 b_{\partial_k}$ 
 & \multicolumn{1}{c|}{\multirow{2}{*}{ $(1,0)\oplus(0,1)$}} &
\multicolumn{1}{c|}{\multirow{2}{*}{$T_2$}}
\\
 & $|\epsilon_{ijk}| \bar{a}_{\partial_k} \gamma_j\gamma_5 b_w-|\epsilon_{ijk}| \bar{a}_w
\gamma_j\gamma_5 b_{\partial_k}$ 
 &  & 
   \\\hline
 
\multicolumn{1}{|c|}{\multirow{4}{*}{$a_2(1,2^{++}$)}} & $Q_{ijk} \bar{a}_{\partial_k} \gamma_j
b_n-Q_{ijk} \bar{a}_n \gamma_j b_{\partial_k}$ 
 & \multicolumn{1}{c|}{\multirow{2}{*}{ $(1,0)\oplus(0,1)$}} & \multicolumn{1}{c|}{\multirow{4}{*}{$E$}}
\\
 & $Q_{ijk} \bar{a}_{\partial_k} \gamma_j b_w-Q_{ijk} \bar{a}_w \gamma_j b_{\partial_k}$ 
 &  & \\\cline{2-3}
  &$Q_{ijk} \bar{a}_{\partial_k} \gamma_j\gamma_t b_n-Q_{ijk} \bar{a}_n \gamma_j\gamma_t
b_{\partial_k}$ 
 & \multicolumn{1}{c|}{\multirow{2}{*}{ $(1/2,1/2)_b$}} & 
 \\
 & $Q_{ijk} \bar{a}_{\partial_k} \gamma_j\gamma_t b_w-Q_{ijk} \bar{a}_w \gamma_j\gamma_t
b_{\partial_k}$
 &  & 
 \\\cline{2-4}
& $|\epsilon_{ijk}| \bar{a}_{\partial_k} \gamma_j b_n-|\epsilon_{ijk}| \bar{a}_n \gamma_j
b_{\partial_k}$ 
 & \multicolumn{1}{c|}{\multirow{2}{*}{ $(1,0)\oplus(0,1)$}} &
\multicolumn{1}{c|}{\multirow{4}{*}{$T_2$}} 
\\
 & $|\epsilon_{ijk}| \bar{a}_{\partial_k} \gamma_j b_w-|\epsilon_{ijk}| \bar{a}_w \gamma_j
b_{\partial_k}$ 
 &  & \\\cline{2-3}
  & $|\epsilon_{ijk}| \bar{a}_{\partial_k} \gamma_j\gamma_t b_n-|\epsilon_{ijk}| \bar{a}_n
\gamma_j\gamma_t b_{\partial_k}$ 
 & \multicolumn{1}{c|}{\multirow{2}{*}{ $(1/2,1/2)_b$}} & 
 \\
 & $|\epsilon_{ijk}| \bar{a}_{\partial_k} \gamma_j\gamma_t b_w-|\epsilon_{ijk}| \bar{a}_w
\gamma_j\gamma_t b_{\partial_k}$
 &  & 
 \\\hline

\multicolumn{1}{|c|}{\multirow{4}{*}{$\pi_2(1,2^{-+}$)}} & $Q_{ijk} \bar{a}_{\partial_k}
 \gamma_j\gamma_t\gamma_5 b_n-Q_{ijk} \bar{a}_n \gamma_j\gamma_t\gamma_5 b_{\partial_k}$ 
 & \multicolumn{1}{c|}{\multirow{2}{*}{ $(1/2,1/2)_a$}} & \multicolumn{1}{c|}{\multirow{2}{*}{$E$}}
\\
 & $Q_{ijk} \bar{a}_{\partial_k} \gamma_j\gamma_t\gamma_5 b_w-Q_{ijk} \bar{a}_w
\gamma_j\gamma_t\gamma_5 b_{\partial_k}$ 
 &  & \\\cline{2-4}
  & $|\epsilon_{ijk}| \bar{a}_{\partial_k} \gamma_j\gamma_t\gamma_5 b_n-|\epsilon_{ijk}| \bar{a}_n
\gamma_j\gamma_t\gamma_5 b_{\partial_k}$ 
 & \multicolumn{1}{c|}{\multirow{2}{*}{ $(1/2,1/2)_a$}} &
\multicolumn{1}{c|}{\multirow{2}{*}{$T_2$}}
\\
 & $|\epsilon_{ijk}| \bar{a}_{\partial_k} \gamma_j\gamma_t\gamma_5 b_w-|\epsilon_{ijk}| \bar{a}_w
\gamma_j\gamma_t\gamma_5 b_{\partial_k}$ 
 &  & 
 \\\hline
\multicolumn{1}{|c|}{\multirow{4}{*}{$\rho(1,1^{--}$)}} & $\bar{a}_{n}
\gamma_j b_n$ 
 & \multicolumn{1}{c|}{\multirow{2}{*}{ $(1,0) \oplus (0,1)$}} & \multicolumn{1}{c|}{\multirow{8}{*}{$T_1$}}
\\
 & $\bar{a}_{w} \gamma_j b_w$ 
 &  & 
 \\\cline{2-3}
  & $\bar{a}_{n}  \gamma_j \gamma_t  b_n$ 
 & \multicolumn{1}{c|}{\multirow{2}{*}{$(1/2,1/2)_b$}} &

\\
 & $\bar{a}_{w}  \gamma_j \gamma_t  b_w$  
 &  & \\
 \cline{1-3}
\multicolumn{1}{|c|}{\multirow{2}{*}{$a_1(1,1^{++}$)}} & $\bar{a}_{n}
\gamma_j \gamma_5 b_n$ 
 & \multicolumn{1}{c|}{\multirow{2}{*}{ $(1,0)\oplus(0,1)$}} & \\
 & $\bar{a}_{w} \gamma_j \gamma_5 b_w$ 
 &  & 
 \\
\cline{1-3}
\multicolumn{1}{|c|}{\multirow{2}{*}{$b_1(1,1^{+-}$)}} & $\bar{a}_{n}
\gamma_j \gamma_t \gamma_5 b_n$ 
 & \multicolumn{1}{c|}{\multirow{2}{*}{ $(1/2,1/2)_a$}} & \\
 & $\bar{a}_{w} \gamma_j \gamma_t \gamma_5 b_w$ 
 &  & 
 \\
\hline
\hline
\end{tabular}
\end{center}
\caption[List of Interpolators]{\sl List of tensor and vector meson interpolators $\mathcal{O}$ classified with respect to the chiral representations $r$ 
and to the irreducible representations $E, T_2$ of the hyper-cubic group $O_{\text{h}}$. The interpolators are defined in accordance with 
Ref.~\cite{Engel:2011aa}; $n$ and $w$ denote two different smearing widths of the quark sources.}
 \label{Table2}
\end{table*} 
   
\subsection{Predictions from $SU(4)$}
\begin{figure}
\centering
\includegraphics[angle=0,width=.85\linewidth]{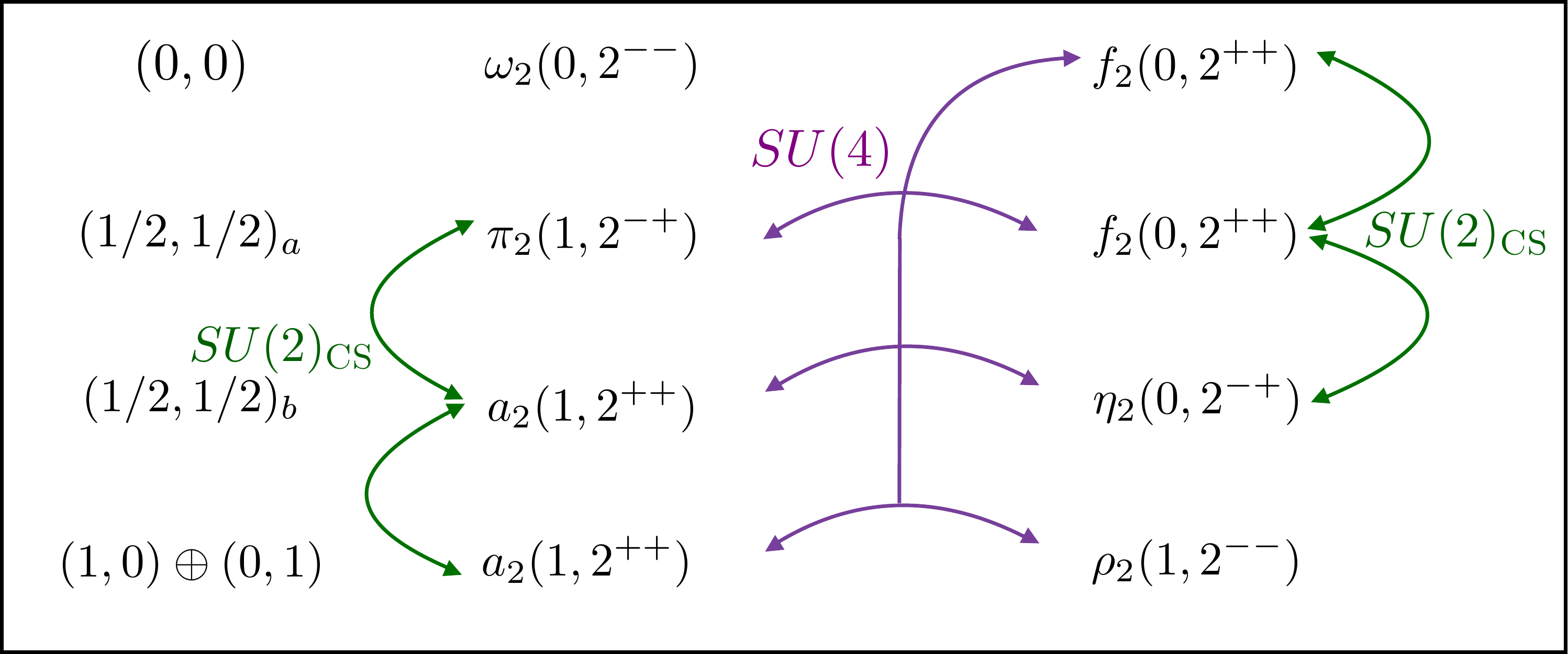}
\caption[SU4 group]{\sl The $SU(2)_{\textsc{cs}}$ triplets are denoted by green lines; $\omega_2$ and $\rho_2$
mesons are $SU(2)_{\textsc{cs}}$ singlets. The $SU(4)$ 15-plet
is indicated by purple lines; $\omega_2$ is the $SU(4)$ singlet.}
 \label{Table3}
\end{figure}

 The states and interpolators from the $(1/2,1/2)_a$ and $(1/2,1/2)_b$ representations have the $\overline{L} R \pm \overline{R} L$ chiral
content, while the states (interpolators) from the $(0,0)$ and $(1,0) \oplus (0,1)$ representations contain the  $\overline{L} L \pm \overline{R} R$ quark combinations.
A symmetry that can connect these representations is the $SU(2)_{\textsc{cs}}$ (\textit{chiralspin}) rotation \cite{Glozman:2014mka, Glozman:2015qva}. The rotations in an
imaginary chiralspin space mix left- and right-handed components of quarks of 
given flavor. The $SU(2)_{\textsc{cs}}$ triplets  and singlets are shown in Fig.~\ref{Table3}.
The $U(1)_{\textsc{A}}$ symmetry is a subgroup of the $SU(2)_{\textsc{cs}}$. 

When we combine both $SU(2)_{\textsc{L}} \times SU(2)_{\textsc{R}}$ and
$SU(2)_{\textsc{cs}}$ symmetries we arrive at a $SU(4)$ group
 with the fundamental vector 
 \begin{align}
\Psi =\begin{pmatrix} u_{\textsc{L}} \\ u_{\textsc{R}}  \\ d_{\textsc{L}}  \\ d_{\textsc{R}} \end{pmatrix} \; , 
\end{align}
 \cite{Glozman:2014mka, Glozman:2015qva}. The $SU(4)$ transformations mix both quarks of different flavors and different chiralities.
 The irreducible representations of $SU(4)$ for $\overline{q} q$ systems are a
 singlet and a 15-plet. The 15-plet includes the following mesons
\begin{align}
 f_2 \longleftrightarrow f_2' \longleftrightarrow \pi_2 \longleftrightarrow a_2 \longleftrightarrow a'_2 \longleftrightarrow \eta_2 \longleftrightarrow \rho_2 \; ,
\end{align}
and the singlet is $\omega_2$, see Fig.~\ref{Table3}. All the states from the 15-plet must be mass-degenerate.
To observe the $SU(4)$ symmetry it is sufficient, however, to see at the same time the degeneracy of one of the chiral multiplets and of one of the 
$SU(2)_{\textsc{cs}}$ triplets. For this purpose it is sufficient to study, e.g.,
all possible isovector mesons $\rho_2, a_2, a_2', \pi_2$.
   
\section{Lattice Technicalities}
\label{Chapter-Lattice-Setup}
 
\subsection{Gauge Field Configurations}
We use two-flavor dynamical Overlap fermion gauge field configurations on a 
$16^3 \times 32$ lattice with lattice spacing $a\sim 0.12 $ fm
generated and generously provided by the JLQCD collaboration, Refs.~\cite{Aoki:2008tq,Aoki:2012pma}. 
The pion mass is $M_{\pi} = 289(2)$ MeV, Ref.~\cite{Noaki:2008iy}. The topological sector is fixed to $Q_{\textsc{T}} = 0$. 
Our gauge ensemble consists of 83 gauge configurations. 

\subsection{Source smearing}
In the previous studies of $J=1$ mesons, Refs.~\cite{Denissenya:2014poa, Denissenya:2014ywa}, we have
used  quark propagators with stochastic sources generated by the
JLQCD collaboration. The spin $J=2$ mesons require
quark propagators with derivatives, however. Given the JLQCD gauge
configurations we calculate the quark propagators using
our standard techniques with different smearing widths of Gaussian 
type \cite{Burch:2004he}. A set of different extended sources with different
smearing widths allows for a larger operator basis in the variational method.

Gaussian smearing has two parameters, the hopping parameter $\kappa$ and the number of smearing steps $N$. It produces Gaussian shaped covariant sources of different widths. 
We choose the same parameters as in Ref.~\cite{Engel:2011aa}, namely ``narrow'' $n$ ($\kappa=0.21, N=18$) and ``wide'' $w$ ($\kappa=0.191, N=41$) sources. The derivative source
is constructed by applying a covariant derivative to the wide source and is denoted as $\partial_k$, with $k=1,2,3$. 

\begin{figure*}[t!]
    \centering 
        \includegraphics[scale=0.55]{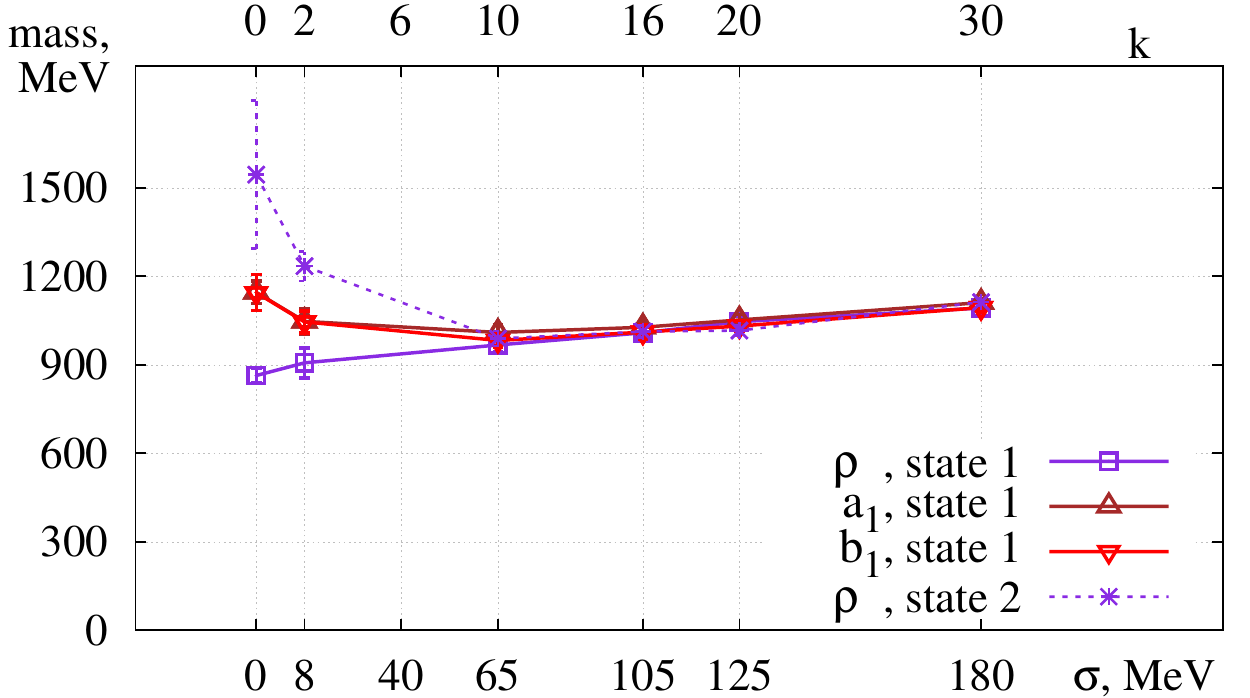} 
      \caption{Mass evolution of $J=1$ isovector mesons.
}\label{fig:j1mass}
   \end{figure*}

   \begin{figure*}[htb]
    \centering 
      \begin{subfigure}[b]{0.45\textwidth}
        \centering
        \includegraphics[scale=0.55]{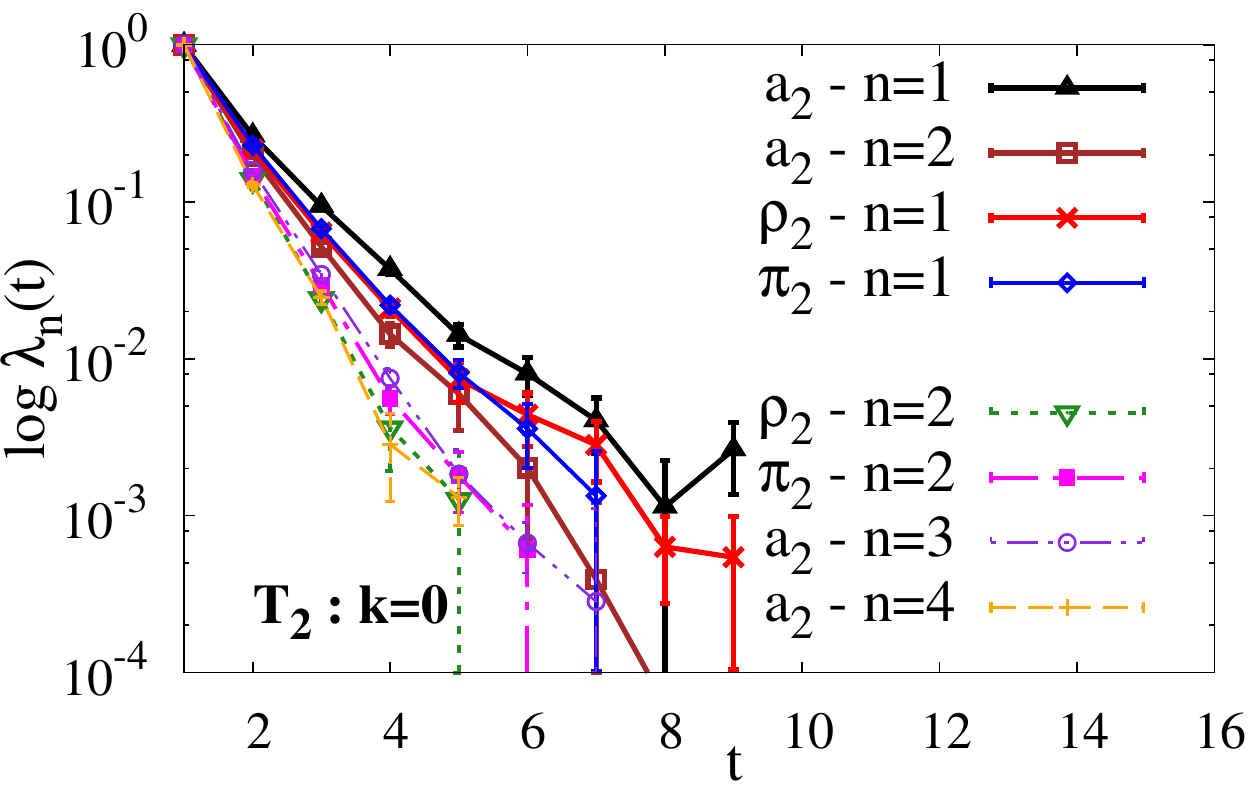}
        \caption{}
      \end{subfigure}
       \hspace*{-32pt}
      \begin{subfigure}[b]{0.45\textwidth}
          \centering
          \includegraphics[scale=0.55]{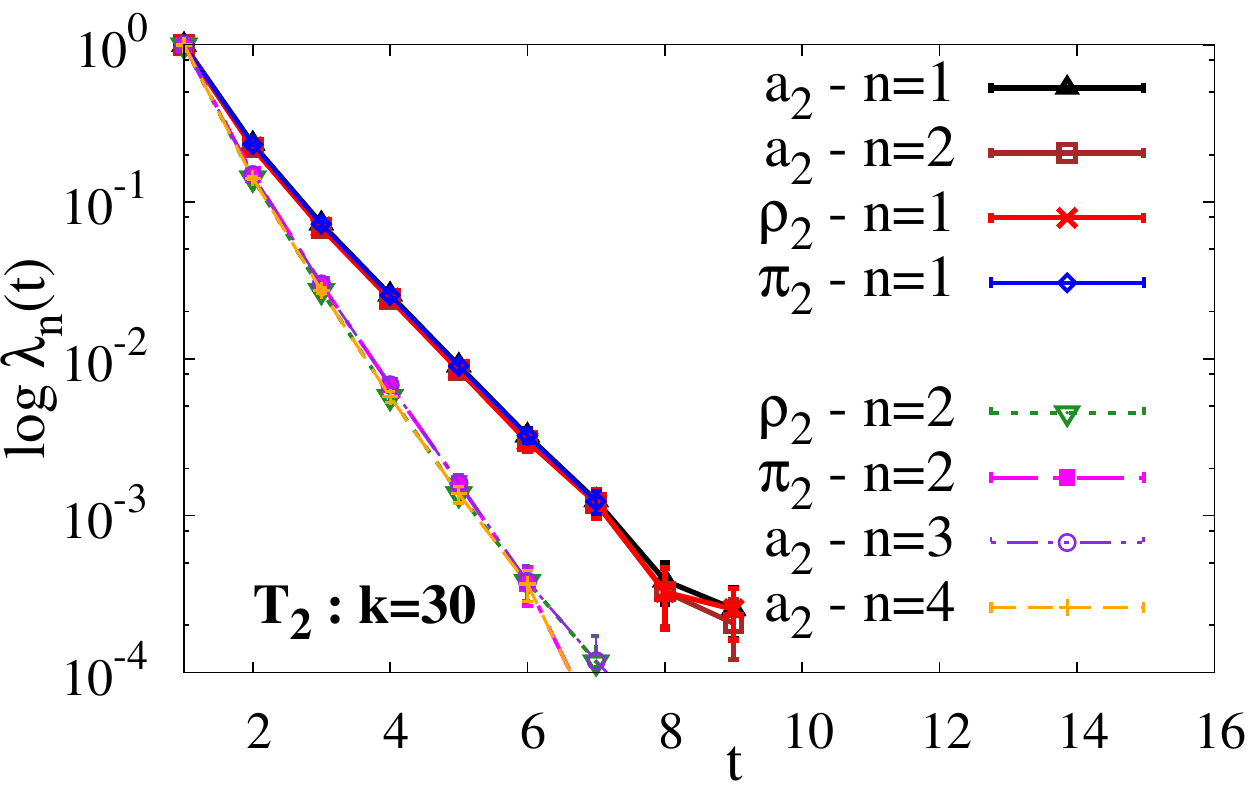}
          \caption{}
      \end{subfigure}
     \hspace*{24pt} \hfill\\
      \caption{Eigenvalues of  $J=2$ tensor mesons: (a) full case ($k=0$), (b) after excluding $k=30$
low modes in $T_2$.
}\label{fig:j2T2ev}
   \end{figure*}
   
\subsection{Truncated Quark Propagator}
We follow the procedure to remove an increasing amount of the lowest-lying
Dirac eigenmodes and study the effect of this reduction on hadron masses.
The truncated quark propagators take the form:
\begin{align}
S_{k}(x,y) = S_{\textsc{full}}(x,y) -  \sum_{i=1}^k \frac{1}{\lambda_i} v_i(x)  v^{\dagger}_i(y) \, .
\end{align}
Here $S_{\textsc{full}}(x,y)$ denotes the untruncated, $S_{k}(x,y)$ the truncated propagator, $\lambda_i$ are the low-lying eigenvalues, 
$v_i(x)$ the eigenvectors and $k$ the number of removed lowest modes. 
For instance, $k=10$ means, that $10$ low modes are removed from the quark propagator. We choose the 
truncation steps in the range $k=2 - 30$, which corresponds to an eigenvalue cutoff between $(8 - 180)$ MeV. 

\subsection{Meson Spectroscopy}
Our analysis is based on the variational
method, see Ref.~\cite{Michael:1985ne, Luscher:1990ck, Blossier:2009kd}. 
The $J=2$ interpolators from Table \ref{Table2} fall 
into the irreducible representation $E$ or $T_2$
of the hyper-cubic group $O_{\text{h}}$\footnote{For a mapping of 
the irreducible representations of $O_{\text{h}}$ to the first few $J$ numbers 
see Ref.~\cite{Johnson1982147}. The interpolators in  $E$ and $T_2$ representations are orthogonal, thus masses can be extracted separately.}.
In addition, our interpolators $\mathcal{O}_i$ fall into different irreducible representations of the parity-chiral and $SU(4)$ groups, as discussed
in the previous chapter.  

We construct the following cross-correlation matrices
\begin{equation}
 C_{ij}(t)= \langle \mathcal{O}_i(t) \mathcal{O}^{\dagger}_j(0)\rangle \; ,
\end{equation}
and solve a generalized eigenvalue problem
\begin{equation}\label{eq:GEVP}
 C(t) \vec{v}_n(t,t_0)= \lambda_n(t,t_0)C(t_0)\vec{v}_n(t,t_0) \; , 
\end{equation}
where $t_0=1$ is chosen as a reference timeslice in our analysis. From the solution of this eigenvalue
problem we determine the energy levels of a given quantum channel. We choose  time ranges where the
eigenvalues $\lambda(t,t_0)$ decay exponentially, i.e.   
\begin{equation}
\lambda^{(n)}(t,t_0) \E^{-E_n (t-t_0)}(1+\mathcal{O}(\E^{-\Delta E_n (t-t_0)}),
\end{equation}
and apply a one-exponential fit to extract masses $E_n$, where
$n$ labels the ground ($n=1$) and excited ($n>1$) states.

\section{ Results}
\label{Chapter-Results}
As a consistency check we first extract masses of the isovector $J=1$
mesons ($\rho, \rho', a_1,b_1$), Fig.~\ref{fig:j1mass}, and compare them to the results shown in Refs.~\cite{Denissenya:2014poa, Denissenya:2014ywa}, 
where propagators obtained within the JLQCD collaboration have been used. This comparison shows
full agreement between both results.
   
\begin{figure*}[htb]
    \centering 
        \begin{subfigure}[b]{0.45\textwidth}
        \centering
        \includegraphics[scale=0.55]{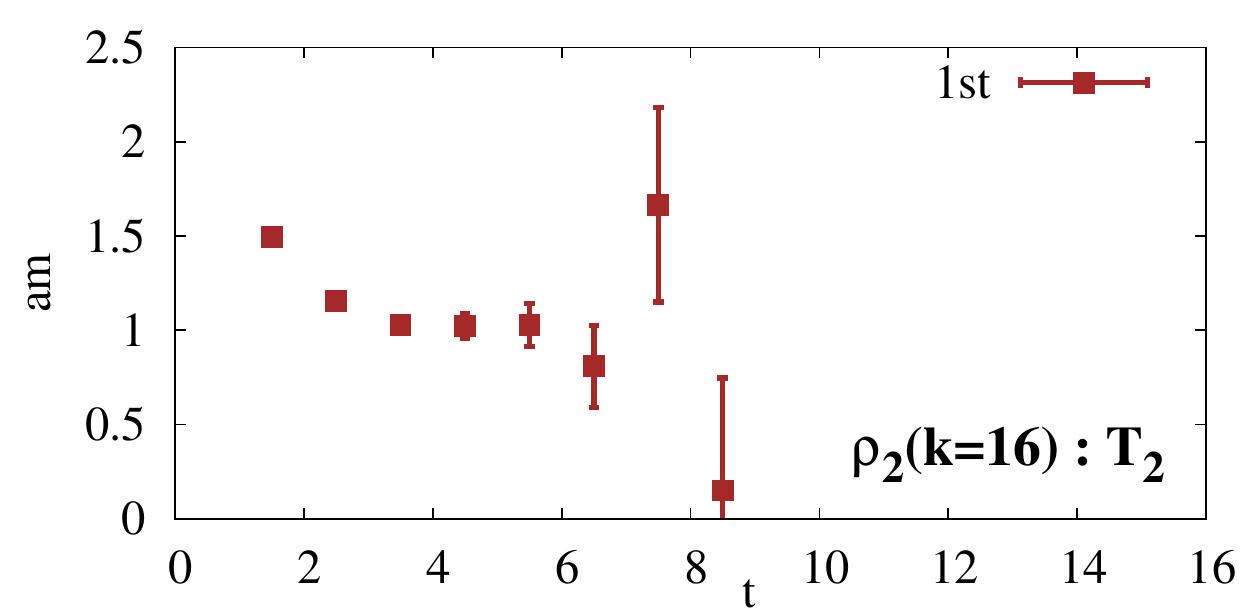}\\
         \includegraphics[scale=0.55]{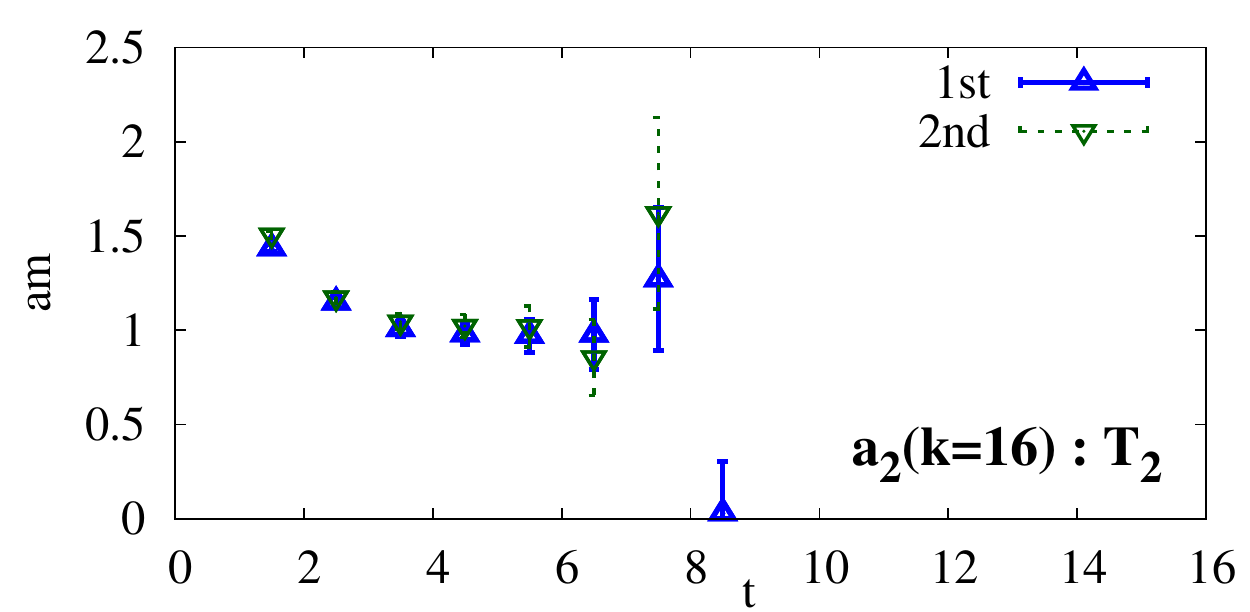}\\   
        \includegraphics[scale=0.55]{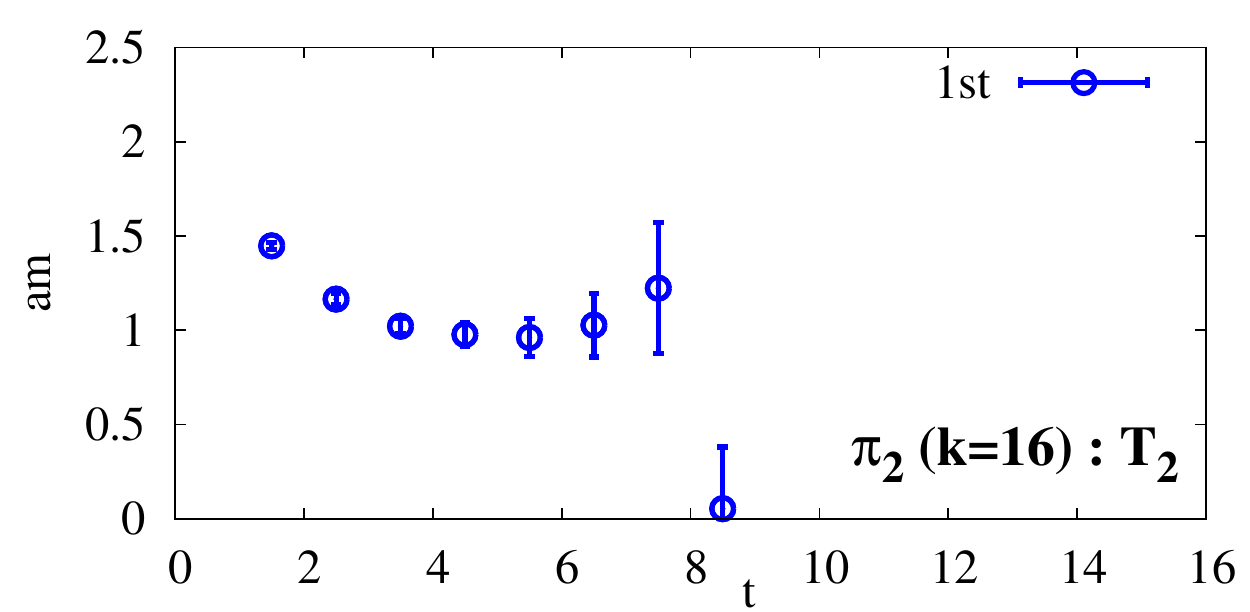}

        \caption{}\label{fig:pi2T2mk0}
      \end{subfigure}
       \hspace*{-32pt}
      \begin{subfigure}[b]{0.45\textwidth}
          \centering
           \includegraphics[scale=0.55]{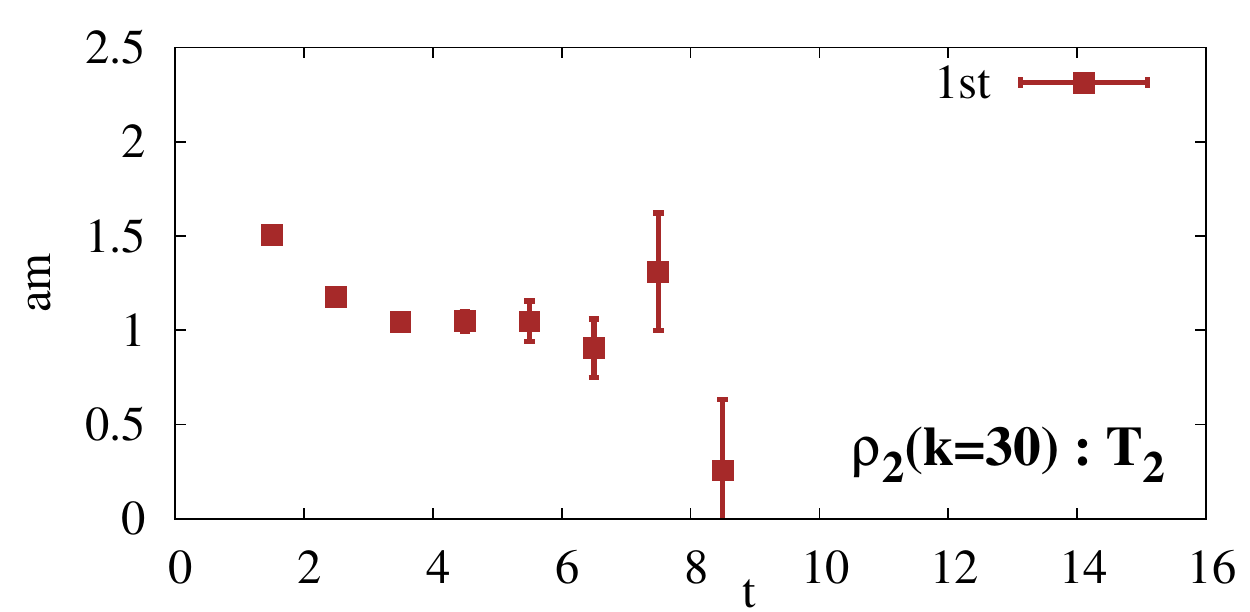}\\
         \includegraphics[scale=0.55]{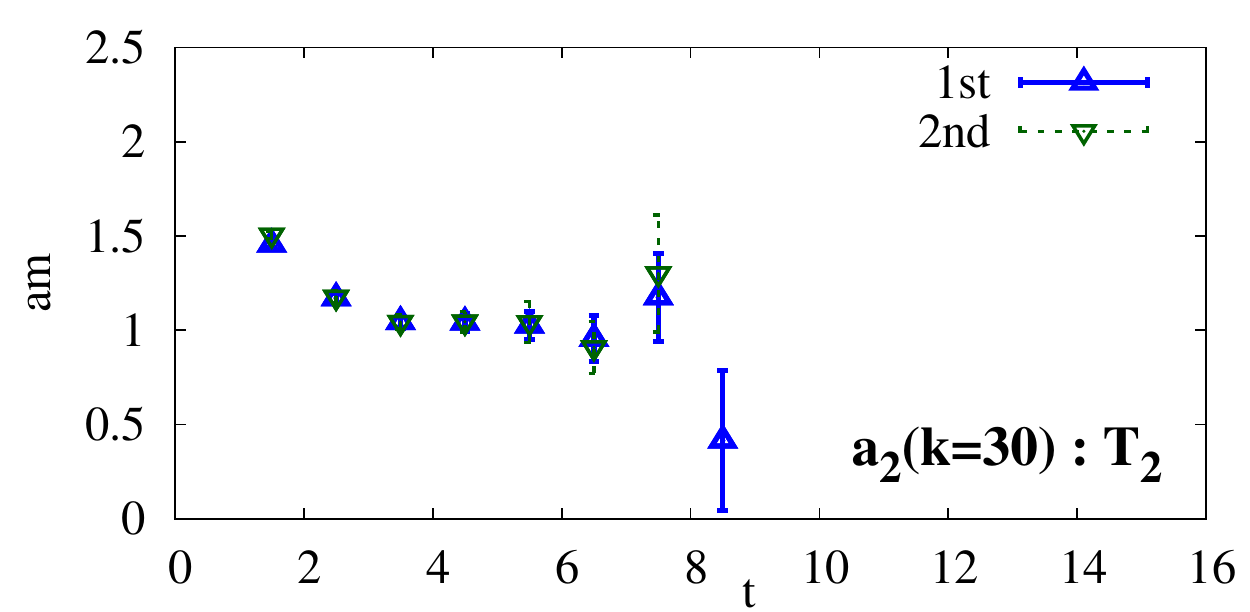}\\       
          \includegraphics[scale=0.55]{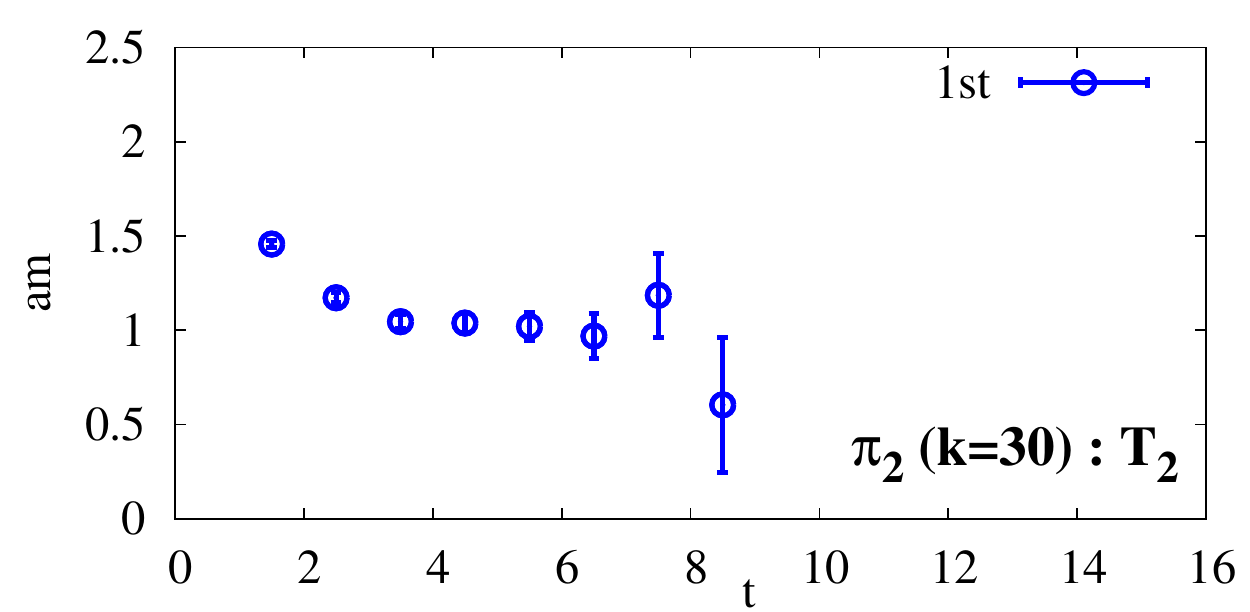}
          \caption{}\label{fig:pi2T2mk30}
      \end{subfigure}
     \hspace*{24pt} \hfill\\
      
            \caption{ $\rho_2$, $a_2$, $a'_2$, $\pi_2$ in $T_2$: Effective masses  after excluding (a) $k=16$, (b)  $k=30$
low modes.
}\label{fig:J2T2m}
   \end{figure*} 
   
\begin{figure*}[htb]
    \centering 
      \begin{subfigure}[b]{0.45\textwidth}
        \centering
        \includegraphics[scale=0.63]{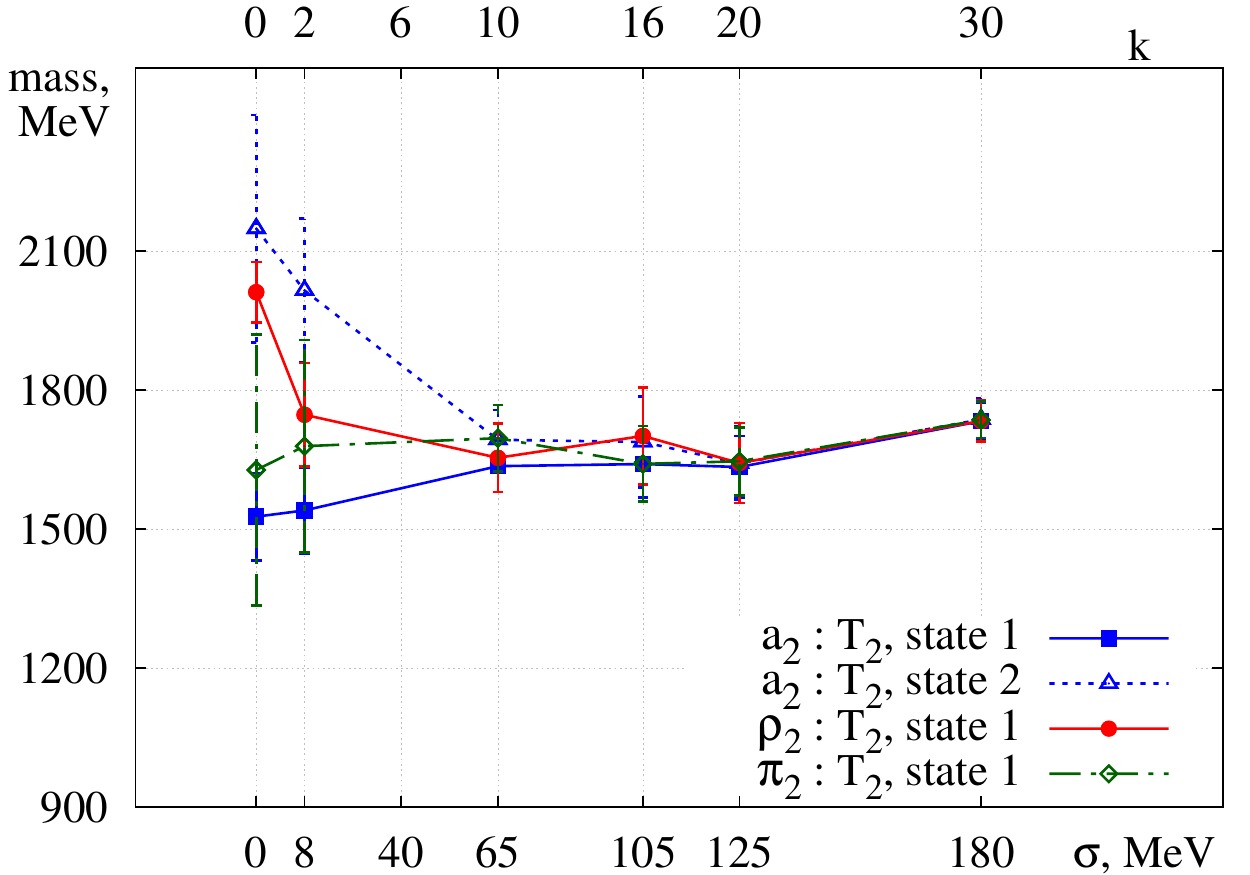}
        \caption{} \label{fig:a}
      \end{subfigure}
       \hspace*{-8pt}
      \begin{subfigure}[b]{0.45\textwidth}
          \centering
          \includegraphics[scale=0.63]{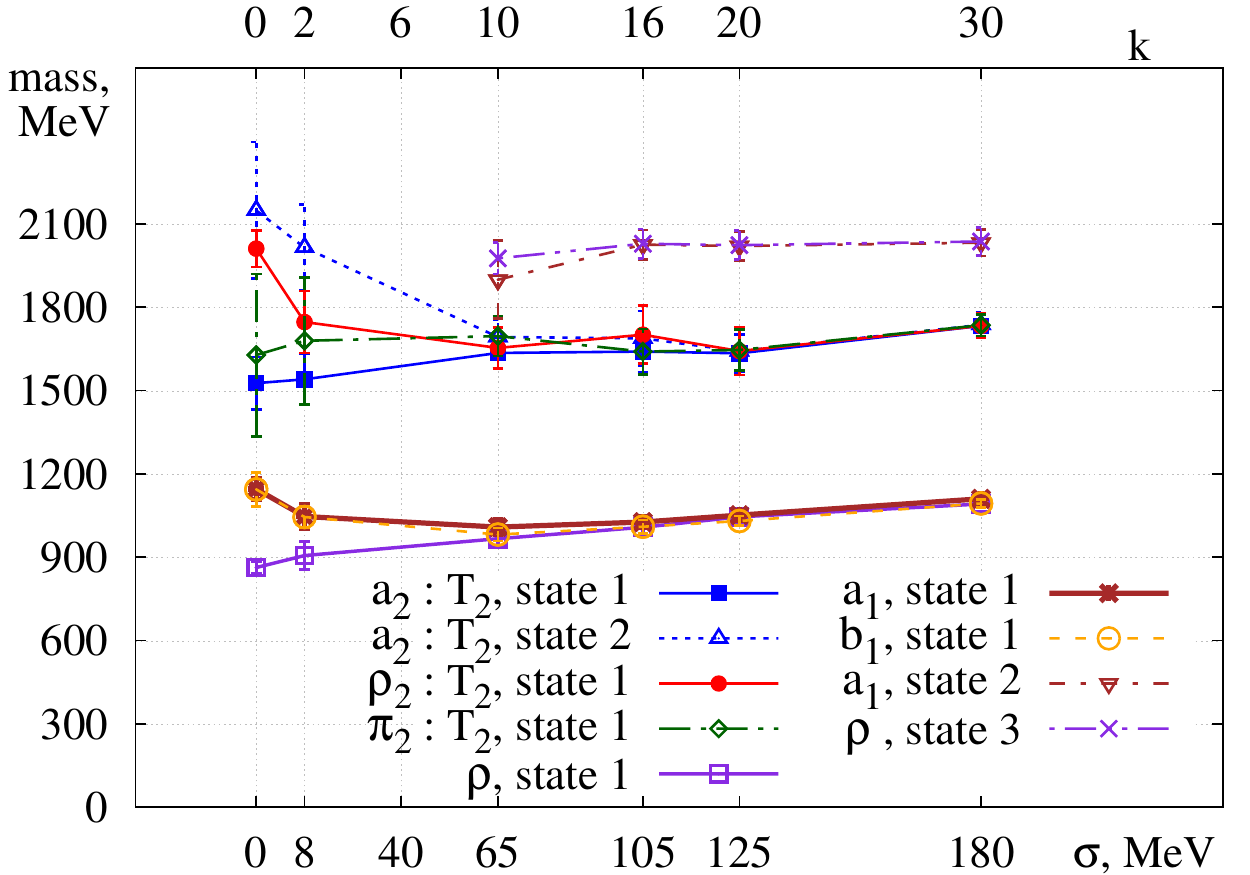}
          \caption{}\label{fig:b}
      \end{subfigure}
     \hspace*{24pt} \hfill\\
      
      \caption{Ground state and excited state mass evolution of (a) $J=2$ mesons in $T_2$ , (b) $J=1$ and $J=2$ mesons. The value $k$ denotes the truncation step and 
$\sigma$ the corresponding energy gap.
}\label{fig:j1j2T2}
   \end{figure*}    

In Fig.~\ref{fig:j2T2ev} we show the eigenvalues of the correlation matrix
for all isovector tensor mesons
before and after removal of the lowest $30$ modes. 
While the eigenvalues are different for the different mesons in the untruncated case, they become identical in the truncated case.
 The degeneracy pattern is completely in line with the expectation from 
the $SU(4)$ symmetry, as discussed in Chapter \ref{Chapter-Chiral-Parity}. Namely,
the $a_2$  ($n=1,2$), $\rho_2$ ($n=1$) and
$\pi_2$ ($n=1$) eigenvalues of the meson cross-correlators in $T_2$ become identical. 
The same holds true for the excited states of $a_2$ ($n=3,4$) $\rho_2$ ($n=2$), $\pi_2$ ($n=2$).  

In Fig.~\ref{fig:J2T2m} we show effective mass plots for all isovector mesons after truncation of the lowest Dirac modes. The quality of the plateau is 
worse than for the $J=1$ states, where no derivative operators are used. This is why the error bars for the extracted masses are larger
than for the $J=1$ case. A comparison to the case of untruncated $J=2$ mesons shows that the quality of the signal essentially improves after truncation, in line with our previous
observations in Refs.~\cite{Denissenya:2014poa, Denissenya:2014ywa,Glozman:2012fj}. The fit ranges and the extracted
masses are given in Appendix \ref{Chapter-Appendix}.

The final results for the masses of the two lowest $a_2$ states and the lowest
states of $\pi_2$ and $\rho_2$ as a function of the truncation level
are given in  Fig.~\ref{fig:a}. We observe a clear onset of the $SU(4)$ symmetry
after elimination of the lowest $10-20$ modes (truncation
energy $65-125$ MeV), in agreement with the respective
results for the $J=1$ isovector mesons shown in Fig.~\ref{fig:j1mass}.

Finally, we compare in Fig.~\ref{fig:b} the lowest energy level for $J=2$ mesons
with the lowest $J=1$ energy level and its first excitation. No degeneracy
is observed between the first excited $J=1$ level and the ground $J=2$ level.
However, to come to a definite conclusion we need much larger volumes: our small
volume could affect in a different manner the $J=1$ radially excited level and
the ground $J=2$ level. 

\section{Summary and Conclusions}
\label{Conclusions}
We have studied  the spin-2 isovector mesons 
upon removal of the lowest-lying Dirac eigenmodes from the valence quark propagators. 
Upon truncation of a small amount of the quasi-zero modes
we have found the same $SU(4)$ symmetry pattern, as in our previous study 
of $J=1$ mesons. This supports the existence of the $SU(4)$ symmetry
in $J \geq 1$ mesons without the quasi-zero modes. This symmetry includes
both the rotations of right- and left-handed quarks in the isospin space
as well as rotations in the chiralspin space that mix the left- and right-handed components. This symmetry group implies the absence of the
color-magnetic field in the system without the quasi-zero modes and might be
interpreted as a manifestation of the dynamical color-electric string in QCD. 

\begin{acknowledgements}
We are grateful to S. Aoki, S. Hashimoto and T. Kaneko for
their suggestion to use the JLQCD overlap gauge configurations and for their hospitality. We are very thankful to C. B. Lang for numerous discussions and help.
Support from the Austrian Science Fund (FWF)
through the grant P26627-N27 is acknowledged. The calculations have been performed on clusters 
at ZID at the University of Graz and 
at the Graz University of Technology.
\end{acknowledgements}

\begin{appendix}
 \section{\label{sec:ff} Masses and Fits}
 \label{Chapter-Appendix}
Single exponential effective mass fits and corresponding $\chi^2$/d.o.f. are presented in Table \ref{Tablek0}
for truncations $k=0,16,20,30$. 

   \begin{table*}[htb]
\begin{center}
\begin{tabular}{lccccclcccccc}
 \hline
 \hline
 \multicolumn{6}{c}{$ k = 0 $} & & \multicolumn{6}{c}{$\ k = 16 $} \\\cline{1-6}\cline{8-13} 
\multicolumn{1}{c}{   state  } &\multicolumn{1}{c}{   $n$  } & $am$   &$\chi^2/d.o.f $ & $t$ & $i$ & &\multicolumn{1}{c}{   state  } &\multicolumn{1}{c}{   $n$  } & $am$   & $\chi^2/d.o.f $ & $t$ &
$i$ \\\hline
\multicolumn{1}{l}{\multirow{2}{*}{$a_2: T_2$}} & 1 & 0.918  $\pm$  0.056  & 6.28/5  & 3 - 9 & \multicolumn{1}{c}{\multirow{2}{*}{2 4 6 8}} & & \multicolumn{1}{l}{\multirow{2}{*}{$a_2: T_2$}} & 1 &
0.986  $\pm$  0.044  & 0.84/3  & 4 - 8 & \multicolumn{1}{c}{\multirow{2}{*}{2 4 6 8}} \\
\multicolumn{1}{c}{ } & 2  & 1.015  $\pm$  0.059  & 0.64/2  & 4 - 7 & & &\multicolumn{1}{c}{ } & 2  & 1.292  $\pm$  0.059  & 0.64/2  & 4 - 7 &\\
\multicolumn{1}{l}{\multirow{1}{*}{$\rho_2 : T_2$}} & 1 & 1.209  $\pm$  0.040  & 1.24/2  & 2 - 5 & \multicolumn{1}{c}{\multirow{1}{*}{2 4}} & & \multicolumn{1}{l}{\multirow{1}{*}{$\rho_2 : T_2$}} & 1
& 1.023  $\pm$  0.063  & 0.82/2  & 4 - 7 & \multicolumn{1}{c}{\multirow{1}{*}{2 4}} \\
\multicolumn{1}{l}{\multirow{1}{*}{$\pi_2 : T_2$}} & 1 & 0.979  $\pm$  0.176  & 0.19/2  & 4 - 7 & \multicolumn{1}{c}{\multirow{1}{*}{6 8}} & & \multicolumn{1}{l}{\multirow{1}{*}{$\pi_2 : T_2$}} & 1 &
0.986  $\pm$  0.049  & 0.69/3  & 4 - 8 & \multicolumn{1}{c}{\multirow{1}{*}{6 8}} \\
\multicolumn{1}{l}{\multirow{1}{*}{$a_2: E$}} & 1 & 0.913  $\pm$  0.078  & 0.37/2  & 4 - 7 & \multicolumn{1}{c}{\multirow{1}{*}{2 8}} & & \multicolumn{1}{l}{\multirow{1}{*}{$a_2: E$}} & 1 & 1.037 
$\pm$  0.039  & 4.02/3  & 4 - 8 & \multicolumn{1}{c}{\multirow{1}{*}{2 8}} \\
\multicolumn{1}{l}{\multirow{1}{*}{$\rho_2 : E$}} & 1 & 1.212  $\pm$  0.034  & 2.80/3  & 2 - 6 & \multicolumn{1}{c}{\multirow{1}{*}{8 10}} & & \multicolumn{1}{l}{\multirow{1}{*}{$\rho_2 : E$}} & 1 &
1.010  $\pm$  0.031  & 1.84/4  & 3 - 8 & \multicolumn{1}{c}{\multirow{1}{*}{8 10}} \\
\multicolumn{1}{l}{\multirow{2}{*}{$\rho \,\,: T_1$}} & 1 & 0.519  $\pm$  0.014  & 1.82/7  & 4 - 12 & \multicolumn{1}{c}{\multirow{2}{*}{1 4 5 8}} & & \multicolumn{1}{l}{\multirow{2}{*}{$\rho \,\,:
T_1$}} & 1 & 0.606  $\pm$  0.014  & 1.38/5  & 6 - 12 & \multicolumn{1}{c}{\multirow{2}{*}{1 4 5 8}} \\
\multicolumn{1}{c}{ } & 2  & 0.609  $\pm$  0.013  & 4.55/7  & 4 - 12 & & &\multicolumn{1}{c}{ } & 2  & 0.928  $\pm$  0.013  & 4.55/7  & 4 - 12 &\\
\multicolumn{1}{l}{\multirow{1}{*}{$a_1 : T_1$}} & 1 & 0.689  $\pm$  0.023  & 4.25/3  & 3 - 7 & \multicolumn{1}{c}{\multirow{1}{*}{1 4}} & & \multicolumn{1}{l}{\multirow{1}{*}{$a_1 : T_1$}} & 1 &
0.618  $\pm$  0.012  & 1.50/4  & 5 - 10 & \multicolumn{1}{c}{\multirow{1}{*}{1 4}} \\
\multicolumn{1}{l}{\multirow{1}{*}{$b_1 \,: T_1$}} & 1 & 0.688  $\pm$  0.036  & 2.11/3  & 3 - 7 & \multicolumn{1}{c}{\multirow{1}{*}{22 25}} & & \multicolumn{1}{l}{\multirow{1}{*}{$b_1 \,: T_1$}} & 1
& 0.607  $\pm$  0.015  & 3.70/5  & 4 - 10 & \multicolumn{1}{c}{\multirow{1}{*}{22 25}} \\
\hline
\hline
 \multicolumn{6}{c}{$ k = 20 $} & & \multicolumn{6}{c}{$\ k = 30 $} \\\cline{1-6}\cline{8-13} 
\multicolumn{1}{c}{   state  } &\multicolumn{1}{c}{   $n$  } & $am$   &$\chi^2/d.o.f $ & $t$ & $i$ & &\multicolumn{1}{c}{   state  } &\multicolumn{1}{c}{   $n$  } & $am$   & $\chi^2/d.o.f $ & $t$ &
$i$ \\\hline
\multicolumn{1}{l}{\multirow{4}{*}{$a_2: T_2$}} & 1 & 0.982  $\pm$  0.040  & 1.04/3  & 4 - 8 & \multicolumn{1}{c}{\multirow{4}{*}{2 4 6 8}} & & \multicolumn{1}{l}{\multirow{4}{*}{$a_2: T_2$}} & 1 &
1.042  $\pm$  0.023  & 0.80/4  & 3 - 8 & \multicolumn{1}{c}{\multirow{4}{*}{2 4 6 8}} \\
\multicolumn{1}{c}{ } & 2  & 1.044  $\pm$  0.028  & 0.93/3  & 3 - 7 & & &\multicolumn{1}{c}{ } & 2  & 0.987  $\pm$  0.028  & 0.93/3  & 3 - 7 &\\
\multicolumn{1}{c}{ } & 3  & 1.459  $\pm$  0.046  & 0.50/3  & 3 - 7 & & &\multicolumn{1}{c}{ } & 3  & 1.456  $\pm$  0.046  & 0.50/3  & 3 - 7 &\\
\multicolumn{1}{c}{ } & 4  & 1.507  $\pm$  0.048  & 1.97/3  & 3 - 7 & & &\multicolumn{1}{c}{ } & 4  & 1.550  $\pm$  0.048  & 1.97/3  & 3 - 7 &\\
\multicolumn{1}{l}{\multirow{2}{*}{$\rho_2 : T_2$}} & 1 & 0.987  $\pm$  0.052  & 3.31/3  & 4 - 8 & \multicolumn{1}{c}{\multirow{2}{*}{2 4}} & & \multicolumn{1}{l}{\multirow{2}{*}{$\rho_2 : T_2$}} & 1
& 1.042  $\pm$  0.026  & 2.00/4  & 3 - 8 & \multicolumn{1}{c}{\multirow{2}{*}{2 4}} \\
\multicolumn{1}{c}{ } & 2  & 1.502  $\pm$  0.052  & 2.17/3  & 3 - 7 & & &\multicolumn{1}{c}{ } & 2  & 1.552  $\pm$  0.052  & 2.17/3  & 3 - 7 &\\
\multicolumn{1}{l}{\multirow{2}{*}{$\pi_2 : T_2$}} & 1 & 0.990  $\pm$  0.044  & 0.62/3  & 4 - 8 & \multicolumn{1}{c}{\multirow{2}{*}{6 8}} & & \multicolumn{1}{l}{\multirow{2}{*}{$\pi_2 : T_2$}} & 1 &
1.043  $\pm$  0.024  & 0.86/4  & 3 - 8 & \multicolumn{1}{c}{\multirow{2}{*}{6 8}} \\
\multicolumn{1}{c}{ } & 2  & 1.477  $\pm$  0.048  & 0.65/3  & 3 - 7 & & &\multicolumn{1}{c}{ } & 2  & 1.467  $\pm$  0.048  & 0.65/3  & 3 - 7 &\\
\multicolumn{1}{l}{\multirow{2}{*}{$a_2: E$}} & 1 & 1.044  $\pm$  0.040  & 9.25/3  & 4 - 8 & \multicolumn{1}{c}{\multirow{2}{*}{2 8}} & & \multicolumn{1}{l}{\multirow{2}{*}{$a_2: E$}} & 1 & 1.053 
$\pm$  0.035  & 5.62/3  & 4 - 8 & \multicolumn{1}{c}{\multirow{2}{*}{2 8}} \\
\multicolumn{1}{c}{ } & 2  & 1.138  $\pm$  0.037  & 3.61/3  & 4 - 8 & & &\multicolumn{1}{c}{ } & 2  & 1.102  $\pm$  0.037  & 3.61/3  & 4 - 8 &\\
\multicolumn{1}{l}{\multirow{2}{*}{$\rho_2 : E$}} & 1 & 1.029  $\pm$  0.031  & 5.62/4  & 3 - 8 & \multicolumn{1}{c}{\multirow{2}{*}{8 10}} & & \multicolumn{1}{l}{\multirow{2}{*}{$\rho_2 : E$}} & 1 &
1.047  $\pm$  0.024  & 8.32/6  & 3 - 10 & \multicolumn{1}{c}{\multirow{2}{*}{8 10}} \\
\multicolumn{1}{c}{ } & 2  & 1.466  $\pm$  0.042  & 4.59/3  & 3 - 7 & & &\multicolumn{1}{c}{ } & 2  & 1.470  $\pm$  0.042  & 4.59/3  & 3 - 7 &\\
\multicolumn{1}{l}{\multirow{4}{*}{$\rho \,\,: T_1$}} & 1 & 0.629  $\pm$  0.011  & 4.87/5  & 5 - 11 & \multicolumn{1}{c}{\multirow{4}{*}{1 4 5 8}} & & \multicolumn{1}{l}{\multirow{4}{*}{$\rho \,\,:
T_1$}} & 1 & 0.657  $\pm$  0.008  & 8.82/6  & 4 - 11 & \multicolumn{1}{c}{\multirow{4}{*}{1 4 5 8}} \\
\multicolumn{1}{c}{ } & 2  & 0.669  $\pm$  0.007  & 8.85/6  & 4 - 11 & & &\multicolumn{1}{c}{ } & 2  & 0.611  $\pm$  0.007  & 8.85/6  & 4 - 11 &\\
\multicolumn{1}{c}{ } & 3  & 1.224  $\pm$  0.030  & 2.55/2  & 3 - 6 & & &\multicolumn{1}{c}{ } & 3  & 1.216  $\pm$  0.030  & 2.55/2  & 3 - 6 &\\
\multicolumn{1}{c}{ } & 4  & 1.245  $\pm$  0.044  & 2.63/2  & 3 - 6 & & &\multicolumn{1}{c}{ } & 4  & 1.278  $\pm$  0.044  & 2.63/2  & 3 - 6 &\\
\multicolumn{1}{l}{\multirow{2}{*}{$a_1 \,: T_1$}} & 1 & 0.632  $\pm$  0.011  & 1.41/5  & 5 - 11 & \multicolumn{1}{c}{\multirow{2}{*}{1 4}} & & \multicolumn{1}{l}{\multirow{2}{*}{$a_1 \,: T_1$}} & 1
& 0.668  $\pm$  0.009  & 4.73/5  & 5 - 11 & \multicolumn{1}{c}{\multirow{2}{*}{1 4}} \\
\multicolumn{1}{c}{ } & 2  & 1.222  $\pm$  0.029  & 2.59/3  & 3 - 7 & & &\multicolumn{1}{c}{ } & 2  & 1.215  $\pm$  0.029  & 2.59/3  & 3 - 7 &\\
\multicolumn{1}{l}{\multirow{2}{*}{$b_1 \,: T_1$}} & 1 & 0.620  $\pm$  0.010  & 11.77/6  & 4 - 11 & \multicolumn{1}{c}{\multirow{2}{*}{22 25}} & & \multicolumn{1}{l}{\multirow{2}{*}{$b_1 \,: T_1$}} &
1 & 0.657  $\pm$  0.009  & 8.38/7  & 4 - 12 & \multicolumn{1}{c}{\multirow{2}{*}{22 25}} \\
\multicolumn{1}{c}{ } & 2  & 1.240  $\pm$  0.045  & 3.94/4  & 3 - 8 & & &\multicolumn{1}{c}{ } & 2  & 1.287  $\pm$  0.045  & 3.94/4  & 3 - 8 &\\
\hline
\hline
\end{tabular}
\end{center}
\caption{Results of fits to the eigenvalues at a truncation level $k = 0,16,20,30$ for $J = 1,2$ mesons. States are denoted by $n = 1, 2,...$. 
Corresponding mass values $am$ are given in lattice units; $t$
denotes the fit range and $i$ labels the interpolators used in the construction of the cross-correlation matrix in a given quantum channel according to 
Ref.~\cite{Engel:2011aa}.}\label{Tablek0}
\end{table*}

\end{appendix}

\end{document}